\documentclass[twocolumn,prX,amsmath,amssymb]{revtex4}

\usepackage{graphicx}
\usepackage{dcolumn}
\usepackage{bm}

\begin{document}

\title{Asymptotic analysis of the $Z$-scan trace to determine \\ multiphoton absorption coefficients}

\author{Carlos J. Zapata-Rodr\'{\i}guez}
\email{carlos.zapata@uv.es}

\address{Departamento de \'Optica, Universidad de Valencia, Dr. Moliner 50, 46100 Burjassot, Spain.}

\date{\today}

\begin{abstract}
We present a theoretical analysis of open aperture $Z$-scan signatures for materials exhibiting an absorption admixture of different multiphoton processes.
Based on a polynomial expansion, we suggest some procedures to separate the contribution to the trace of the recorded transmittance made by absorptions of distinct order.
The signal processing from experimental data allows a sequential estimation of assorted nonlinear absorption coefficients accurately.
\end{abstract}

%\ocis{190.0190, 190.4400.}

\maketitle

\section{\label{sec01} Introduction}
 
Multiphoton absorption of intense wavefields in matter is a phenomenon of interest in a great variety of applications.
In multiphoton microscopy, for example, enhanced optical sectioning is achieved leading to superresolved three-dimensional imaging without the use of a confocal aperture.\cite{Denk90,Gu96b}
In this case, higher order nonlinear absorption (NLA) processes are mostly proposed to further improve spatial resolution and light penetration.\cite{Larson03,Polesana05}
In optical limiting, on other hand, highly nonlinear absorbing materials are also expected to exhibit large transmittance at low intensities, resistance to laser-induced damage, stability over time, and ultrafast responses.\cite{Stryland85,Said97,Venkatram05}

In pursue of characterizing the nonlinear response of a material, open aperture $Z$-scan provides the standards to determine the multiphoton absorption coefficients.\cite{Sheik90}
In this technique, the transmittance of a thin sample is measured in the far field when it is scanned through the focal volume of a focused pulse beam.
Profiting from the characteristic theoretical curve drawn at a given multiphoton absorption process, the corresponding NLA coefficient is subsequently determined by fitting the absorptive $Z$-scan trace.
When three or more photons are involved in the NLA process, however, the open aperture scan by itself is not very sensitive.\cite{Hernandez04,Hernandez04b,Correa07}

This small sensitivity is of special significance when more than one nonlinear process is active.\cite{Shim98}
If a pulse beam induces a NLA effect involving a large number of photons, a lower-order absorption process may come out simultaneously within some narrow spectral windows.
For instance, ultrashort pulse propagation at $1320\ \mathrm{nm}$ in the polydiacetylene-PTS exhibits simultaneous 2PA (two-photon absorption) and 3PA.\cite{Yoshino03}
Also, an exact degeneracy between three- and four-photon absorption is observed at the peak wavelength of $1890\ \mathrm{nm}$.
%Importantly, both NLAs are resonantly enhanced.\cite{Yoshino03}
In chalcogenide glasses, it is found an important contribution of 2PA in measuring the three-photon NLA coefficient at $1.06\ \mu\mathrm{m}$ in the picosecond range.\cite{Boudebs04,Cherukulappurath05}
Even four- and five-photon absorption is observed in $\mathrm{TeO}_2$-based samples at a wavelength of $1550\ \mathrm{nm}$ with a small linear absorption coefficient.\cite{Bindra01} 

In the examples given above, there is no significant difference between the experimental $Z$-scan and the fits obtained using $n$- and $(n+1)$-photon absorption coefficients as the fitting parameter.
To distinguish between these two absorptions, the $Z$-scan experiment is commonly accomplished at various irradiances.\cite{Shim98,Yoshino03,Bindra01,Lawrence94,Ganeev04}
For sufficiently low intensities, the $(n+1)$-photon absorption process cannot participate in the $Z$-scan signature; on the contrary this effect clearly dominates over the absorptance peak of the curve at high irradiances.
Nevertheless, one single $Z$-scan trace measured at an intermediate irradiance might provide the required NLA coefficients simultaneously.
Pointing upon this direction, analytical or semi-analytical approximations\cite{Gu05} are proposed to simplify this labour.

In this paper we address the problem of identifying and separating the contributions of degenerate multiphoton absorption processes from a single $Z$-scan signature. 
Our analysis exploits the fact that $Z$-scan traces yield accurate estimates of NLA coefficients if the curve fitting is performed with experimental data taken with sufficient precision.
The paper is organized as follows.
In Sec.~\ref{sec02} we review the basic grounds on multiphoton absorption occurring in a thin slab made of a nonlinear material.
Analytical expressions of open aperture $Z$-scan signatures for Gaussian pulse beams are deduced in Sec.~\ref{sec03}.
In Sec.~\ref{sec04} we interpret the $Z$-scan traces conveniently and we give some procedures to achieve accurate estimates for the order of the nonlinearity and its corresponding absorption coefficient of the different NLA processes.
Finally, in Sec.~\ref{sec05} the main conclusions are outlined.

\section{\label{sec02} Nonlinear absorption by a thin medium}

Let us consider a pulsed beam propagating in air along the $z$ axis and impinging normally onto a thin bulk medium to prove its response upon NLA.
%The thin slab of material has a width $L$.
In this study we assume that if the beam intensity is low enough to neglect NLA in the sample, the transverse profile is not disturbed significantly due to diffraction.
Exceptionally, linear absorption in the thin sample is regarded.
Additionally, we consider ultrashort pulse excitations so that slow (cumulative) nonlinearities such as thermal nonlinearity are also ignored.\cite{Shinkawa08,Ogusu08}
Thus we examine $m$-photon absorption and $l$-photon absorption simultaneously, being $2 \le m, l$.
Starting from the slowly varying envelope approximation we may simplify the evaluation of the field observed at the exit plane by means of the following differential equation\cite{Stryland98}
\begin{equation}
 \partial_{\zeta_{m}} \tilde{I} = - \tilde{I}^{m} - \eta_{m l} \tilde{I}^{l}
\label{eq05h}
\end{equation}
and the contour condition $\tilde{I} = 1$ at $\zeta_{m} = 0$.
We point out that a linear absorption process may be included in Eq.~(\ref{eq05h}) if $l = 1$ (or $m = 1$) and, alternatively, if we insert the term $- \eta_{m 1} \tilde{I}$ on the right-hand side of the equation, cases that are considered below.
Here, $\tilde{I} = I / I_0$ stands for the ratio of the output intensity $I$ and the input intensity $I_0$ in the sample, the reduced axial coordinate $\zeta_m = L / L_m$ is the ratio of the sample depth $L$ and the characteristic length for the $m$-photon absorption process,
\begin{equation}
 L_m = \alpha^{-1}_m I_0^{1-m} ,
\label{eq05a}
\end{equation}
being $\alpha_m$ the $m$-photon nonlinear absorption coefficient.
%For chalcogenide glasses, typically,\cite{Boudebs04} $\alpha_2 = 8\ \mathrm{cm}/\mathrm{GW}$ and $\alpha_3 = 0.1\ \mathrm{cm}^3/\mathrm{GW}^2$ so that $L_2 = 0.067\ \mathrm{cm}$ and $L_3 = 1.1\ \mathrm{cm}$ for a peak intensity $I_0 = 3\ \mathrm{GW}/\mathrm{cm}^2$. 
For instance, in polydiacetylenes typically\cite{Yoshino03} $\alpha_2 = 6\ \mathrm{cm}/\mathrm{GW}$ and $\alpha_3 = 2.7\ \mathrm{cm}^3/\mathrm{GW}^2$ at $1320\ \mathrm{nm}$ so that $L_2 = 830\ \mu\mathrm{m}$ and $L_3 = 930\ \mu\mathrm{m}$ for a peak intensity $I_0 = 2\ \mathrm{GW}/\mathrm{cm}^2$. 
Finally,
\begin{equation}
 \eta_{m l} = \frac{L_{m}}{L_{l}} = \frac{\alpha_l}{\alpha_m}  I_0^{l-m}
\label{eq05h2}
\end{equation}
represents the relative strength of both NLA processes.
In our numerical example $\eta_{2 3} = 0.9$ suggesting a balanced charge of both nonlinear phenomena.
%For the sake of completeness let us point out that 2PA would dominate for a peak intensity $I_0 = 0.2\ \mathrm{GW}/\mathrm{cm}^2$ where $\eta_{2 3} = 0.09$ and, contrarily, 3PA would be stronger at higher intensities; for instance $\eta_{2 3} = 9.0$ at $I_0 = 20\ \mathrm{GW}/\mathrm{cm}^2$.

The permutation $m \leftrightarrow l$ in (\ref{eq05h}) provides an equivalent form of the differential equation; if one combination shows $\eta_{m l} \ge 1$ the counterpart gives $\eta_{l m} = \eta_{m l}^{-1} \le 1$.
We may restrict our problem considering $m$PA a dominant process over $l$PA, whence leaving the unity as the maximum value of the coupling parameter $\eta_{m l}$.
According to Eq.~(\ref{eq05h2}), $\eta_{m l} \le 1$ at sufficiently low intensities if $m < l$ . 

Validity of Eq.~(\ref{eq05h}) is restricted to nonlinear interactions with matter producing small and modest alterations of the beam profile.
Otherwise, multiphoton absorption may lead to self-focusing within the sample.
In mathematical terms we impose $\zeta_{m} |\partial_{\zeta_{m}} \tilde{I}| \ll \tilde{I}$.
Using Eqs.~(\ref{eq05a}) and (\ref{eq05h2}) we infer a sufficient condition $ L \ll L_n $ (or $\zeta_n \ll 1$) for $n = m,l$.

In general, one cannot find an analytical solution of Eq.~(\ref{eq05h}).
For a single NLA process being $\eta_{m l} = 0$, a solution may be provided in a closed form.
In this case, Eq.~(\ref{eq05h}) is reduced to $\partial_{\zeta_m} \tilde{I} = - \tilde{I}^m$, and the exit intensity in the medium yields
\begin{equation}
 \tilde{I} = \frac{1}{\sqrt[m-1]{1 + (m-1) \zeta_m}} .
\label{eq05b}
\end{equation} 
%\begin{equation}
% I (r,z+L,t) = \frac{I (r,z,t)}{\sqrt[m-1]{1 + (m-1) L \alpha_m I^{m-1} (r,z,t)}}
%\label{eq05b}
%\end{equation} 
In some circumstances, e.g. $(m-1) \zeta_m \ge 1$, Eq.~(\ref{eq05b}) provides a significant beam depletion due to strong $m$-photon absorption, a result that is out of the validity in our approach.
Therefore we focus our attention on the regime $\zeta_m \ll 1$.
In this case we may expand Eq.~(\ref{eq05b}) into Taylor series around $\zeta_m = 0$ giving
\begin{eqnarray}
 \tilde{I}^{(S)} = 1 + \sum_{s=1}^S \frac{\zeta_m^s}{s!} \prod_{p=0}^{s-1} [p-pm-1] + O[\zeta_m]^{S+1} ,
\label{eq05c}
\end{eqnarray}
%\begin{eqnarray}
% I^{(S)} (r,z+L,t) &=& I (r,z,t) \\
%             &+& \sum_{s=1}^S \prod_{p=0}^{s-1} [p-pm-1] \frac{1}{s!} L^s \alpha_m^s I^{sm - s +1} (r,z,t) \nonumber
%\label{eq05c}
%\end{eqnarray}
where $S$ is a positive integer.
In the limit $S \to \infty$, $I^{(\infty)}$ provides an exact solution of the transmitted field intensity. 
We point out that $\zeta_m = 0$ stands for sufficiently thin samples but also for sufficiently low input intensities such that NLA of $m$ photons is negligible, in accordance to the definition $\zeta_m = \alpha_m L I_0^{m-1}$.
Specifically $\tilde{I}^{(1)} = 1 - \zeta_m$, and the parabolic approximation yields
\begin{eqnarray}
 \tilde{I}^{(2)} = \tilde{I}^{(1)} + \frac{m}{2} \zeta_m^2 = 1 - \zeta_m + \frac{m}{2} \zeta_m^2, \ (\eta_{m l} = 0).
\label{eq05d}
\end{eqnarray}

Interestingly, we may provide a solution of the more complex equation (\ref{eq05h}) also in terms of polynomials of $\zeta_m$.
We point out that a different approach using the method of Adomian may provide an approximate solution of (\ref{eq05h}) given also in terms of a series expansion.\cite{Sanchez00,Sanchez06}
Developing $\tilde{I}$ into a Taylor series about the origin, the first-order solution yields
%\begin{eqnarray}
% I^{(1)} (r,z+L,t) &=& I (r,z,t) \\
%                   &-& \alpha_{m_1} L I^{m_1} (r,z,t) - \alpha_{m_2} L I^{m_2} (r,z,t) \nonumber
%\label{eq05g}
%\end{eqnarray}
%equivalently,
\begin{equation}
 \tilde{I}^{(1)} = \tilde{I}_0 + (\partial_{\zeta_m} \tilde{I})_0 \zeta_m = 1 - (1 + \eta_{m l}) \zeta_{m} ,
\label{eq05i}
\end{equation}
where subindex $0$ stands for the specific value at $\zeta_m = 0$.
Eqs.~(\ref{eq05c}) for $S=1$ [also shown in (\ref{eq05d})] and (\ref{eq05i}) are formally equivalent within this linear approach; however, for simultaneous NLA processes we use $(1 + \eta_{m l}) \zeta_{m}$ instead of $\zeta_{m}$.
In this context we interpret that, in a low-order approximation, our solution for competing NLAs might correspond to the solution for a single one with $\zeta_m \to L / L_{m l}$ and characteristic absorption depth
\begin{equation}
 L_{m l} = \frac{L_m}{1 + \eta_{m l}} = \frac{L_m L_l}{L_m + L_l} .
\label{eq05j}
\end{equation}
The average length $L_{m l}$ approaches $L_m$ in the low-intensity regime where $L_m \ll L_l$, but shifts toward $L_l$ as long as the input intensity increases.
In general, since two NLA processes are present, $L_{m l}$ is lower than either $L_m$ or $L_l$ individually.
The quadratic (second-order) approximation obtained from the Taylor series yields
%\begin{eqnarray}
% I^{(2)} (r,z+L,t) &=& I^{(1)} (r,z+L,t) \\
%                   &+& \frac{m_1 \alpha_{m_1}^2}{2} L^2 I^{2 m_1 - 1} (r,z,t) \nonumber \\
%                   &+& \frac{m_2 \alpha_{m_2}^2}{2} L^2 I^{2 m_2 - 1} (r,z,t) \nonumber \\
%                   &+& \frac{(m_1 + m_2) \alpha_{m_1} \alpha_{m_2}}{2} L^2 I^{m_1 + m_2 - 1} (r,z,t) \nonumber
%\label{eq05g}
%\end{eqnarray}
%equivalently
\begin{eqnarray}
 \tilde{I}^{(2)} = \tilde{I}^{(1)} + \frac{(1 + \eta_{m l}) (m + l \eta_{m l})}{2} \zeta_{m}^2 .
\label{eq05k}
\end{eqnarray}
We point out that the last term of the polynomial expansion is evaluated using the chain rule 
\begin{equation}
 \partial_{\zeta_m}^2 \tilde{I} = \partial_{\tilde{I}} (\partial_{\zeta_m} \tilde{I}) \partial_{\zeta_m} \tilde{I} 
\end{equation}
and Eq.~(\ref{eq05h}).
Once again, Eq.~(\ref{eq05k}) would be associated with beam depletion under a single NLA process [see (\ref{eq05d})] in using the axial coordinate $(1 + \eta_{m l}) \zeta_{m}$, as previously discussed within the linear approximation, and also substituting the number $m$ of photons by the average
\begin{equation}
 \bar{m} = \frac{m + l \eta_{m l}}{1 + \eta_{m l}} .
\label{eq05l}
\end{equation}
This parameter spans from $m$ to $l$ as $\eta_{m l}$ moves from zero to infinity ($\eta_{l m} \to 0$); at the intermediate value $\eta_{m l} = 1$, $\bar{m}$ stands for the arithmetic mean of $m$ and $l$.

\begin{figure}
\centering
\includegraphics[width=7.5cm]{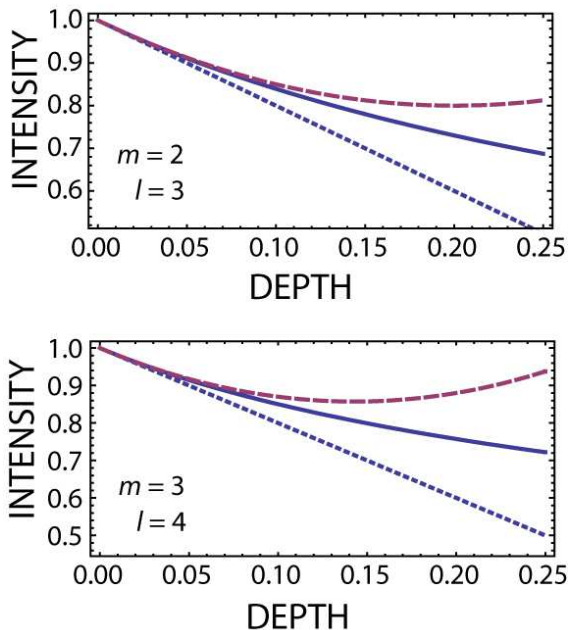}
\caption{Evolution of $\tilde{I}$ when $m$PA and $l$PA are present simultaneously.
Depth $\zeta_m$ is in the horizontal axis and the parameter $\eta_{m l} = 1$.
Numerical evaluation of (\ref{eq05h}) is plotted in solid lines.
Also, the linear approximation (\ref{eq05i}) is drawn in dotted lines, and $I^{(2)}$ of Eq.~(\ref{eq05k}) is represented in dashed lines.} 
\label{fig01}
\end{figure}

Solution of Eq.~(\ref{eq05h}) is computed numerically and represented graphically in Fig.~\ref{fig01} for different values of $m$ and $l$.
A unit value of $\eta_{m l}$ is selected in order to analyze NLAs with balanced strengths within the sample.
The linear approach given in (\ref{eq05i}) overestimates field attenuation due to absorption, whereas the quadratic approximation (\ref{eq05k}) underestimates it.
The validity of these approximations is restricted to low depths, $\zeta_m \ll 1$, and deviations upon the exact solution become severe if $\zeta_m$ approaches unity.
Though not represented in the figure, convergence with higher-order solutions results extremely slow showing strong oscillations upon the order $S$ of the approach if $\zeta_m \lesssim 1$.
This suggests that a Fourier expansion in terms or harmonic functions could be more convenient than a Taylor series of polynomials.
Nevertheless we follow a different approach next. 

Let us exploit the fact that beam depletion under simultaneous $m$PA and $l$PA behaves similarly to single $\bar{m}$PA, for which the exact solution (\ref{eq05b}) might be provided.
Hence, using the axial coordinate $(1 + \eta_{m l}) \zeta_{m}$ instead of $\zeta_{m}$ in Eq.~(\ref{eq05b}), and substituting the number $m$ of photons absorbed in the process by the average $\bar{m}$ given in (\ref{eq05l}), we finally have
\begin{equation}
 \tilde{I} \approx \frac{1}{\sqrt[\bar{m}-1]{1 + (\bar{m}-1) (1 + \eta_{m l}) \zeta_m}} .
\label{eq05m}
\end{equation} 
In the limit $\eta_{m l} = 0$, obviously, Eq.~(\ref{eq05m}) simplifies to Eq.~(\ref{eq05b}) giving the exact solution.

\begin{figure}
\centering
\includegraphics[width=7.5cm]{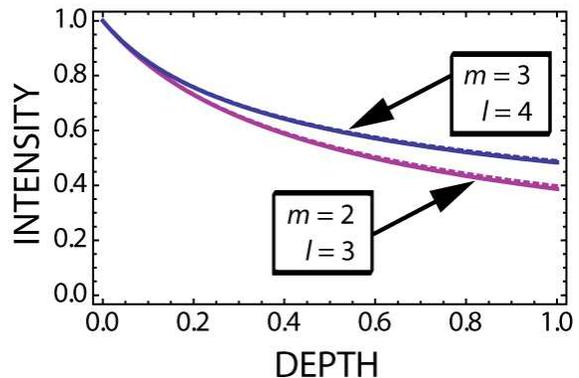}
\caption{Behavior of $\tilde{I}$ within depths $\zeta_m \le 1$, assuming $m$PA and $l$PA occur at $\eta_{m l} = 1$.
Numerical evaluation of (\ref{eq05h}) is plotted in solid lines.
Also, the analytical approach given in (\ref{eq05m}) is drawn in dotted lines.} 
\label{fig02}
\end{figure}

In Fig.~\ref{fig02} we inspect graphically the accuracy of (\ref{eq05m}) for the intermediate value $\eta_{m l} = 1$.
Within the interval of interest, $0 \le \zeta_m \le 1$, deviations of the computed intensities and those available from Eq.~(\ref{eq05m}) have the upper bound $9.7\ 10^{-3}$ found at $\zeta_2 = 1$ for mixed 2PA and 3PA processes, representing a relative error of less than $1\%$.
For simultaneous 3PA and 4PA, accuracy improves since deviation decreases to $5.1\ 10^{-3}$ at $\zeta_3 = 1$.

\subsection{Samples with strong linear absorption}

Literature values for the linear-optical absorption coefficient $\alpha_1$ in the near-infrared region vary between $0.1$ and $100\ \mathrm{cm}^{-1}$ for PTS,\cite{Krug89} the example given above, giving characteristic lengths as short as $L_1 = 100\ \mu\mathrm{m}$.
Linear absorption dominates over NLA processes at sufficiently low values of $I_0$ and, therefore, such an effect cannot be neglected in this regime.
In this sense, a solution of Eq.~(\ref{eq05h}) may be provided for $l = 1$, that is, when linear absorption and only a single nonlinear process is active, which is written as \cite{Cherukulappurath05,Hermann84}
\begin{equation}
 \tilde{I} = \frac{\exp \left( - \eta_{m 1} \zeta_m \right)}{\sqrt[m-1]{1 + (m-1) \zeta_{\mathrm{eff} m}}} .
\label{eq05n}
\end{equation}
Here, 
\begin{equation}
 \zeta_{\mathrm{eff} m} = \frac{1 - \exp \left[ - (m - 1) \eta_{m 1} \zeta_m \right]}{(m - 1) \eta_{m 1}} 
\label{eq05o}
\end{equation}
stands for the effective axial coordinate.
Importantly, the product $\eta_{m 1} \zeta_m$ ($ \triangleq \zeta_1$) is independent of $L_m$.
Also, $\zeta_{\mathrm{eff} m}$ and Eq.~(\ref{eq05n}) lead to $\zeta_m$ and Eq.~(\ref{eq05b}), respectively, if $\eta_{m 1} = 0$.

Analyticity of the solution (\ref{eq05n}) suggests that an approach of $\tilde{I}$ may be given for the complete equation
\begin{equation}
 \partial_{\zeta_{m}} \tilde{I} = - \eta_{m 1} \tilde{I} - \tilde{I}^{m} - \eta_{m l} \tilde{I}^{l} ,
\label{eq05p}
\end{equation}
which accounts for linear absorption and two different NLA processes, simultaneously.
Accordingly, we may follow the procedure given above identifying the terms in the parabolic series expansion of $\tilde{I}$ about the origin, $\zeta_m = 0$, derived here from Eq.~(\ref{eq05n}) and also from Eq.~(\ref{eq05p}).
Using over Eq.~(\ref{eq05n}) the normalization of the axial coordinate $\zeta_m \to L / L_{m l}$ in terms of the average length of Eq.~(\ref{eq05j}), and also substituting the number $m$ of photons by the average $\bar{m}$ of (\ref{eq05l}), lead to severe simplifications. 
The required approach is straightforwardly attained if we additionally substitute $\eta_{m 1} = L_m / L_1$ by 
\begin{equation}
 \frac{L_{m l}}{L_1} = \frac{\eta_{m 1}}{1 + \eta_{m l}} ,
\label{eq05q}
\end{equation}  
giving
\begin{equation}
 \tilde{I} = \frac{\exp \left( - \eta_{m 1} \zeta_m \right)}{\sqrt[\bar{m}-1]{1 + (\bar{m}-1) (1 + \eta_{m l}) \zeta_{\mathrm{eff} \bar{m}}}} .
\label{eq05r}
\end{equation}
The effective axial coordinate $\zeta_{\mathrm{eff} \bar{m}}$ is also conveniently transformed from (\ref{eq05o}) by using the average $\bar{m}$ instead of the integer $m$.
%\begin{equation}
% \zeta_{\mathrm{eff} \bar{m}} = \frac{1 - \exp \left[ - (\bar{m} - 1) \eta_{m 1} \zeta_m \right]}{(\bar{m} - 1) \eta_{m 1}} ,
%\label{eq05s}
%\end{equation}

\begin{figure}
\centering
\includegraphics[width=7.5cm]{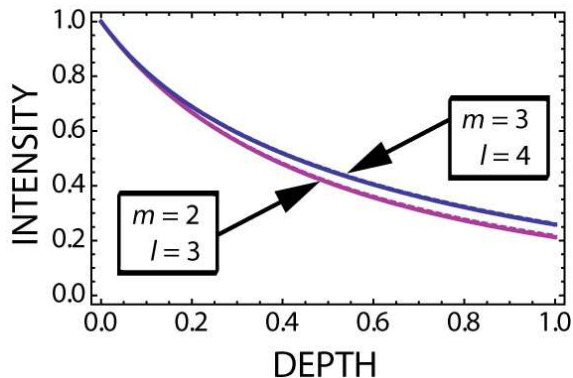}
\caption{Dependence of $\tilde{I}$ at different depths when $m$PA, $l$PA, and linear absorption are found simultaneously at $\eta_{m l} = 1$ and $\eta_{m 1} = 1$.
Again, the numerical estimation of $\tilde{I}$ is plotted in solid lines, and the approximation given in (\ref{eq05r}) is drawn in dotted lines.} 
\label{fig03}
\end{figure}

Validity of Eq.~(\ref{eq05r}) is examined in Fig.~\ref{fig03} for balanced absorption processes, i.e., $\eta_{m l} = 1$ and $\eta_{m 1} = 1$.
Numerical simulations are also depicted, and an excellent agreement with Eq.~(\ref{eq05r}) is remarkable.
We point out that the approach (\ref{eq05r}) provides the exact value of $\tilde{I}$ in the limits $\eta_{m l} = 0$, $\eta_{m 1} = 0$, $\eta_{m l} \to \infty$, and $\eta_{m 1} \to \infty$.

\section{\label{sec03} $Z$-scan signatures with Gaussian beams}

\begin{figure*}
\centering
\includegraphics[width=15cm]{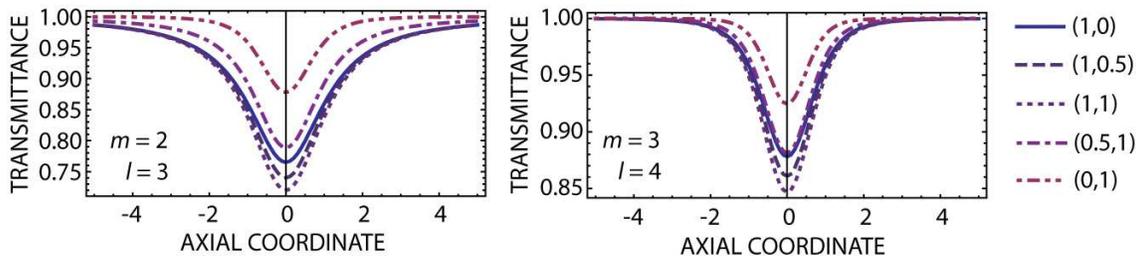}
\caption{$Z$-scan signatures for  2PA and 3PA admixes (left) and simultaneous 3PA and 4PA processes (right).
Dominant $m$PA with $\Delta_m = 1$ (and $\Delta_l = 0$) is drawn in solid line.
Other values of the ordered pair $(\Delta_m,\Delta_l)$ are represented in different curve styles.} 
\label{fig04}
\end{figure*}

From an experimental point of view, nonlinear samples are commonly proved with paraxial Gaussian beams focused with low-numerical-aperture lenses.
Top-hat beams and trimmed Airy beams are claimed to present some advantages over Gaussian beams for $Z$-scan experiments\cite{Rhee96}; however, beam shaping of the laser beam is required.
Also, Gaussian beams are found more attractive in theoretical studies because the three-dimensional distribution of the field is an analytic solution of the paraxial wave equation.\cite{Siegman86}
In this study we assume a sufficiently narrow bandwidth of the pulsed wave in order to neglect spatiotemporal couplings associated with ultrafast beam propagation.\cite{Porras98,Zapata07}
For convenience, we use normalized spatial coordinates: the radial coordinate $r$ involves a normalization with respect to the beam waist $w_0$ and the axial coordinate $z$ is given in units of the Rayleigh range $z_R = k w_0^2 / 2$ being $k = \omega / c$ the wavenumber in air.
If the waist plane of the Gaussian beam is found at $z = 0$ (neglecting focal shifts,\cite{Li81} this is the focal plane of the lens), the instantaneous intensity at the input plane of the scanned sample yields
\begin{equation}
 I_0 (r,z,t) = E(t) w^{-2} \exp \left( - 2 r^2 / w^2 \right) ,
\label{eq02}
\end{equation}
being 
\begin{equation}
 w(z) = \sqrt{1 + z^2}
\label{eq03}
\end{equation}
the Gaussian width, and $E(t)$ the intensity of the pulse envelope at focus, $(r,z) = (0,0)$.
%For a rectangular pulse we would employ $E(t) = E_0$ at $|t| \le 1/2$ ---$E(t) = 0$ otherwise.---
The temporal coordinate $t$ supports a normalization over the pulse duration such that
\begin{equation}
  \frac{1}{E_0} \int_{-\infty}^\infty dt E(t) = 1 ,
\label{eq09}
\end{equation}
where $E_0$ is the maximum instantaneous intensity.
For the sake of convenience, from hereon we consider pulse waveforms with a Gaussian profile.
In this case we write $E(t) = E_0 \exp (- \pi t^2)$ so that its FWHM yields $\Delta \tau = 2 \sqrt{\ln{2} / \pi} \approx 0.94$.
We point out that our analysis may be extended to other waveforms straightforwardly.

In the open-aperture mode of $Z$-scan measurements, light transmitted through the sample is fully collected.
Mismatching of the linear index of refraction of the sample and air induces the incident beam is partially reflected, an effect that may be treated on the experimental data and is therefore ignored here.
The output intensity $I = I_0 \tilde {I}$ is obtained from Eqs.~(\ref{eq05r}) and (\ref{eq02}), which varies in space and time.
For a given pulse, the coordinate $\zeta_m$ reaches a maximum value 
\begin{equation}
 \Delta_m = \alpha_m L E_0^{m-1} 
\label{eq15}
\end{equation}
at focus.
Also the coupling parameter satisfies $\eta_{m l} \le \Delta_{l} / \Delta_{m}$ (if $m < l$).
Upon recording of the integrated intensity at the exit plane of the scanned sample,
\begin{eqnarray}
 P(z) = \int_{-\infty}^\infty dt \int_0^\infty 2 \pi r dr I(r,z,t) ,
\label{eq07}
\end{eqnarray}
changes are found in the vininity of the Gaussian focus, $|z| < 1$.
In the far-field, $|z| \gg 1$, the output intensity $I \to I_0 \exp \left( -\Delta_1 \right)$ so that Eq.~(\ref{eq07}) yields
\begin{eqnarray}
 P_\infty = \lim_{z \to \infty} P(z) = \frac{\pi}{2} E_0 \exp \left( -\Delta_1 \right) .
\label{eq07b}
\end{eqnarray}
Commonly, $Z$-scan signatures refer to traces of the ratio $T = P / P_\infty$, which is interpreted as the normalized transmittance of the sample as being scanned axially.

In order to simplify our discussion remaining in Sec.~\ref{sec03}, let us neglect linear absorption by using the limit $\Delta_1 \to 0$.
In Fig.~\ref{fig04} we depict $Z$-scan signatures when the sample experiences 2PA and 3PA simultaneously (subfigure on the left), and when 3PA and 4PA are present (on the right).
The strength of each NLA process is characterized by the corresponding parameters $\Delta_m$ and $\Delta_l$.
Numerical simulations of the transmittance $T$ are performed using the approximated equation (\ref{eq05m}); we have observed by direct comparison with numerical computation of Eq.~(\ref{eq05h}) that these traces are of extremely-high accuracy for low and intermediate values of $\Delta_m$ and $\Delta_l$.
$Z$-scan responses are symmetric even functions with respect to the origin, $z = 0$.
As expected, NLA is maximum at the Gaussian focus.
The V-shaped response $T$ shows a distinct tail decay and valley value, a fact that may be employed to determine NLA coefficients of the sample from experimental data.
However, a procedure may be given to overcome the inherent lack of precision in the data fit when simultaneous NLA are present, specially as the number of absorbed photons in each nonlinear process increases.

We may take advantage of Eqs.~(\ref{eq05i}) and (\ref{eq05k}) in order to obtain analytical approaches of $Z$-scan signatures.
After denormalizing $\tilde{I}^{(1)}$ conveniently and inserting into Eq.~(\ref{eq07}), we finally have a first-order $T$ response,
\begin{equation}
 T^{(1)} = 1 - \frac{\Delta_m}{w^{2 (m - 1)} \sqrt{m^3}} - \frac{\Delta_l}{w^{2 (l - 1)} \sqrt{l^3}} .
\label{eq08}
\end{equation}
Using $\tilde{I}^{(2)}$ instead, a second-order approach of the $Z$-scan signature yields
\begin{eqnarray}
\label{eq09}
 T^{(2)} &=& T^{(1)} + \frac{(l+m) \Delta_l \Delta_m}{2 w^{2 (m+l-2)} \sqrt{(m+l-1)^3}} \\
               &+& \frac{l \Delta_l^2}{2 w^{4 (l-1)} \sqrt{(2 l - 1)^3}} + \frac{m \Delta_m^2}{2 w^{4 (m-1)} \sqrt{(2 m - 1)^3}} . \nonumber
\end{eqnarray}
Again, in Eq.~(\ref{eq02}) we have considered a bell-shaped Gaussian pulse.
Following a similar approach, Eric W. Van Stryland \emph{et al.}\cite{Sheik90} have provided an infinite series for a solitary 2PA process ($m = 2$ and $\Delta_l = 0$).
For simultaneous $m$PA and $l$PA we might give also an expansion of infinite number of constituents; however, the most relevant terms are provided in $T^{(2)}$.
We point out that top-hat pulses (and of some other shapes) yield also analytical expressions for the coefficients of the series expansion (\ref{eq09}).

\begin{figure}
\centering
\includegraphics[width=7.5cm]{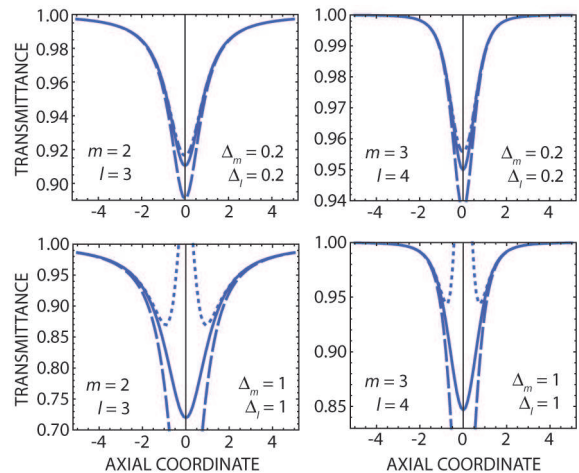}
\caption{$T(z)$ response evaluated numerically (solid curves) and from the linear approach (\ref{eq08}), $T^{(1)}$, in dashed lines and the quadratic approximation (\ref{eq09}), $T^{(2)}$, in dotted lines.}
%Values of parameters $\Delta_m = \Delta_l = 0.2$ (top) and $\Delta_m = \Delta_l = 1$ (bottom). 
\label{fig05}
\end{figure}

High accuracy of $T^{(2)}$ is observed at $\Delta_l,\Delta_m < 10^{-1}$; this fact is expected also from numerical computations depicted in Fig.~\ref{fig01}.
In these cases, however, the collected signal slightly deviates from the unit transmittance associated with the response in the linear regime.
This effect is noticeable at increasing integers $m$ and $l$ where NLA is highly localized in space (around the focus) and time (about the local time $t = 0$).

Values of $\Delta_m$ and $\Delta_l$ commonly exceed $10^{-1}$ in empirical demonstrations of $Z$-scans, so we may test the performance of Eqs.~(\ref{eq08}) and (\ref{eq09}) in those practical cases.
We tentatively assume that deviations of the normalized transmittance $T$ induced by self-focusing are negligible.
In Fig.~\ref{fig05} we plot $Z$-scan signatures for mixed $m$PA and $l$PA of different orders, however keeping a balanced peak depth $\Delta_m = \Delta_l$.
A modest approach of $T^{(1)}$ and $T^{(2)}$  is observed for a value $\Delta_m = 2\ 10^{-1}$; discrepancies are significant only at the Gaussian focus and nearest axial points.
When $\Delta_m$ reaches the unity, the series expansion $T^{(S)}$ manifests a strongly-oscillatory convergence at increasing $S$, specifically in the focal region of the Gaussian beam.
In all cases, however, an excellent agreement of $T^{(1)}$ and $T^{(2)}$ responses is found along the signature tails.

\section{\label{sec04} Determination of the NLA coefficients}

In spectroscopic measurements, absorption coefficients are determined after data processing by fitting the response $T$ within the region of interest; commonly this is the focal region of the Gaussian beam where the signal changes stronger.
The detected transmittance valley has a shape and a value at the bottom that may help to determine NLA coefficients; however, this method has intrinsically poor sensibility for two or more unknowns and may lead to ill estimates.
On other hand, far-field decay of $Z$-scan signatures, which is delineated accurately in Eqs.~(\ref{eq08}) and (\ref{eq09}), reveal distinctively the NLA orders of the absorptions involved and the values of their respective coefficients.
As a consequence, data fitting may be carried out alternatively onto points of the trace out of focus, assuming that weak disturbances of the transmittance may be discerned in spite of fluctuations.
%Let us next give a thorough analysis exploiting this idea.

\begin{figure}
\centering
\includegraphics[width=7.5cm]{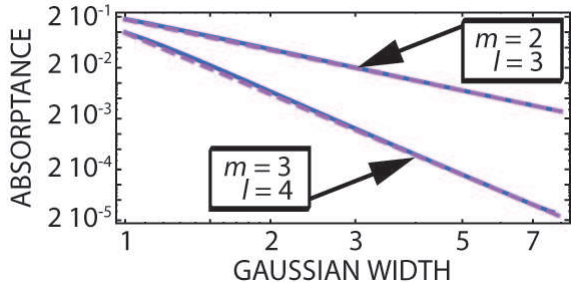}
\caption{Asymptotic behaviour of the absorptance $A$ versus the Gaussian width $w$, drawn in solid lines.
Balanced depths $\Delta_m = \Delta_l = 1/2$ have been used in the simulations.
Dashed lines correspond to asymptotes (\ref{eq10}).}
\label{fig06}
\end{figure}

Instead of using the normalized transmittance $T$ experienced in the sample as being scanned along the focal region, for convenience, our theoretical analysis is simplified in terms of the absorptance 
\begin{equation}
 A(z) = 1 - T(z)
\end{equation}
produced in the nonlinear material.
Specifically, the curve $A(z)$ traces a centered peak of symmetric vanishing tails.
%From a fundamental point of view, the $T$ response stands for the field transmittance .
%Alternatively, $1 - T$ would represent the absorptance on the medium providing a centered peak of symmetric vanishing tails.
In accordance with (\ref{eq08}) and (\ref{eq09}), and assuming that $m < l$, the absorptance tails mainly fall in direct proportion to $w$ raised to the power of $-2 (m - 1)$.
For illustration, in Fig.~\ref{fig06} we show a log-log plot of the peak $A$ versus the Gaussian width $w$.
In the far field, $w \gg 1$, the absorptance draws a straight line
\begin{equation}
  \log{ A } = - 2 (m - 1) \log{w} + \log{\left[\frac{\Delta_m}{\sqrt{m^3}}\right]} ,
\label{eq10}
\end{equation}
given in the slope-intercept form.
We point out that discrepancies of $A$ and its asymptote (\ref{eq10}) are not significant even at $w \gtrsim 1$ for small and moderate values of $\Delta_l$ in comparison with $\Delta_m$.
This asymptote has a negative slope of absolute value $2 (m - 1)$, thus the integer $m$ may be revealed unambiguously by linear curve fitting to experimental data.
The intercept of the function (\ref{eq10}) found at $w = 1$ displays $m^{- 3/2} \Delta_m$ and subsequently the peak depth may be estimated.
Ultimately the $m$PA coefficient $\alpha_m$ is determined from Eq.~(\ref{eq15}) if the sample width $L$ and peak intensity $E_0$ are known.

\begin{figure}
\centering
\includegraphics[width=7.5cm]{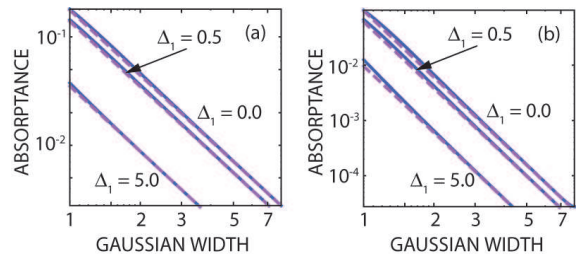}
\caption{Absorptance $A$ (solid lines) and its asymptotic curve (dashed lines) for opaque samples ($\Delta_1 \neq 0$) in the linear regime.
Values $\Delta_m = \Delta_l = 1/2$ are employed in the simulations for: (a) $m = 2$ and $l = 3$, and (b) $m = 3$ and $l = 4$.}
\label{fig07}
\end{figure}

Linear absorption is being omitted in the geometric analysis of the absorptance.
In Fig.~\ref{fig07} we investigate the asymptotic behaviour of $A$ beyond the Gaussian focus for different values of $\Delta_1 = \alpha_1 L$.
In comparison with the asymptote of Eq.~(\ref{eq10}) valid for $\Delta_1 = 0$, opacity of the sample (in the linear regime) leads to a parallel asymptote; thus its slope is still characteristic of the order $m$ of the nonlinearity.
However, its intercept at $\omega = 1$ that allows an estimate of $\Delta_m$ is undervalued.
This is not surprising since the parameter $\Delta_m$ stands for the maximum value of $\zeta_m$, which is a coordinate that leaves out of consideration linear absorption.
Otherwise, we have suggested above the use of the effective axial coordinate $\zeta_{\mathrm{eff} m}$ given in Eq.~(\ref{eq05o}) instead.
In accordance we define the effective parameter $\Delta_{\mathrm{eff} m} = \varrho_m \Delta_m$ as the maximum value of $\zeta_{\mathrm{eff} m}$, where
\begin{equation}
 \varrho_m = \frac{1 - \exp \left[ - (m - 1) \Delta_1 \right]}{(m - 1) \Delta_1}
\end{equation}
denotes the rate decrease of the estimate $\Delta_m$, and therefore the estimate decrease of $\alpha_m$, in the presence of linear absorption.
Fig.~\ref{fig07} also shows the corresponding asymptotes of Eq.~(\ref{eq10}) when $\Delta_m$ is substituted by $\Delta_{\mathrm{eff} m}$, which is in agreement with our analysis.

The steps developed above may be applied trivially for samples experiencing only one NLA process ($\Delta_l = 0$).
To the knowledge of the author, this is the first time a procedure is proposed to derive $\alpha_m$ asymptotically from a single $Z$-scan trace.
Interestingly, in Refs.~\onlinecite{Sheik90}, \onlinecite{Correa07}, and \onlinecite{Tykwinski02}, alternate asymptotic analyses are proposed for single NLAs based on absorptances measured at the Gaussian focus for different peak intensities $E_0$ of the input beam.
Using $w = 1$ and $\Delta_m = \alpha_m L E_0^{m-1}$, Eq.~(\ref{eq10}) leads to 
\begin{equation}
  \log{A} = (m-1) \log{E_0} + \log{\left[\frac{\alpha_m L}{\sqrt{m^3}}\right]} .
\label{eq11}
\end{equation}
Once again, the order $m$ and the coefficient $\alpha_m$ may be determined geometrically in a log-log representation of $A$ versus $E_0$ from the slope and intercept of the asymptote (\ref{eq11}), respectively.
If linear absorption cannot be neglected, we would find a multiply factor $\varrho_m$ to correct the value of $\alpha_m$.
In the case that two NLA processes are manifested simultaneously, the conclusions given above are already valid assuming that $m$ is the lowest order of the nonlinearity.

\begin{figure}
\centering
\includegraphics[width=7.5cm]{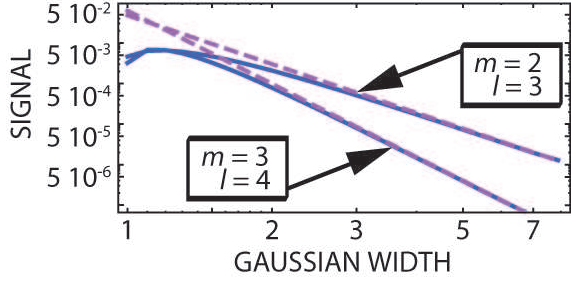}
\caption{Signal $A - \Delta_m / w^{2 (m - 1)} m^{3/2}$ in solid lines and asymptote (\ref{eq12}) in dashed lines for $\Delta_m = \Delta_l = 1/2$.}
\label{fig08}
\end{figure}

Following with light-matter interactions involving an admix of two different NLA processes, we emphasize that accurate measurements leading to a precise $Z$-scan trace should contain sufficient information to extract not only the order $m$ and the NLA coefficient $\alpha_m$ but, additionally, estimates of the parameters $l$ and $\alpha_l$. 
For that purpose, we propose to substract the foremost term of $T^{(1)}$ (apart from unity), $\Delta_m / w^{2 (m - 1)} \sqrt{m^3}$, yet considered, from the absorptance peak $A$.
The tail of the resultant signal has a characteristic decay, faster than the removed part, which depends on the remaining NLA competing in the sample.
In Fig.~\ref{fig08} we illustrate this procedure with some numerical examples.
In the far field, $w \to \infty$, the signal exhibits an asymptote
\begin{eqnarray}
\label{eq12}
 \log{\left[A -  \frac{\Delta_m}{w^{2 (m - 1)} \sqrt{m^3}}\right]} = - 2 (l - 1) \log{w} \nonumber \\
  + \log{\left[\frac{\Delta_l}{\sqrt{l^3}} - \frac{m \Delta_m^2 \delta_{2 m} \delta_{3 l}}{2 \sqrt{(2 m - 1)^3}}\right]} ,
\end{eqnarray}
where $\delta$ denotes the Kronecker's delta.
In a log-log plot, the slope of (\ref{eq12}) displays the order $l$ of the second nonlinear process and the intercept includes the value of the peak depth $\Delta_l$.
In the particular case $m=2$ and $l=3$, however, the intercept incloses a term depending on the squared of $\Delta_m$.
This is not a major problem since $\Delta_2$ is estimated beforehand.
We point out that if also $\Delta_2 = \Delta_3 = 1$, which by the way are values in the boundaries of validity of our approach, then both terms are equivalent in magnitude but having opposite signs and they cancel each other out.
As a consequence, higher-order terms of the series expansion of $T$ would dominate the asymptotic behavior of the signal.

From the numerical simulations shown in Fig.~\ref{fig08} we are concerned about the extreme accuracy of the transmittance data, at least in the order of $10^{-3}$--$10^{-4}$, required to draw the trace of the signal in the asymptotic regime.
Unfortunately, experimental measurements commonly lacks of sufficient precision to apply this method.
In these cases, data fit is suitably performed along segments of the optical axis where absorptance is higher, for instance, at the Gaussian focus.
At this point, additionally, the contribution of the higher-order NLA to the absorptance peak, denoted as $A_{max}$, is enhanced.
Asymptotic analyses of $A_{max} - \Delta_m / \sqrt{m^3}$ in terms of $E_0$, in the way carried out in Eqs.~(\ref{eq11}) and (\ref{eq12}) at $w = 1$, are available in order to estimate $l$ and $\Delta_l$, but would demand also a strong agreement with the experimental data.

\begin{figure}
\centering
\includegraphics[width=7.5cm]{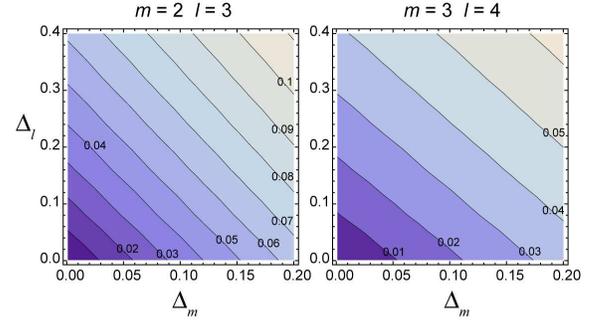}
\caption{Normalized peak absorptance $A_{max}$ for different parameters $\Delta_m$ and $\Delta_l$ and different NLA orders.}
\label{fig09}
\end{figure}

Assuming that the order $l$ of the nonlinear process is known, the peak depth $\Delta_l$ and thus the NLA coefficient $\alpha_l$ is preferably determined by giving the best fit of $A_{max}$ to data.
In Fig.~\ref{fig09} we depict contour plots of the peak absorptance $A_{max}$ for degenerate NLA of different orders in terms of the peak depths.
The plots show a monotonic smooth variation allowing the estimation of $\Delta_l$, which represents the only unknown remaining in our problem, by means of a data fit.

Theoretical evaluation of $A_{max}$ displayed in Fig.~\ref{fig09} is carried out numerically since we cannot provide a simple analytical expression of the absorptance at $w = 1$.
However a series expansion may be obtained from $T^{(S)}$ as $S \to \infty$, which yields
\begin{eqnarray}
 A_{max} &=& \sum_{p=1}^\infty \frac{C_{p} (m)}{p!} \Delta_m^p + \sum_{q=1}^\infty \frac{C_{q} (l)}{q!} \Delta_l^q            \nonumber \\
         &+& \sum_{p,q=1}^\infty \frac{C_{p q} (m,l)}{(p+q)!} \Delta_m^p \Delta_l^q .
\label{eq17}
\end{eqnarray}
Inserting Eq.~(\ref{eq05c}) into Eq.~(\ref{eq07}) leads to  
\begin{eqnarray}
 C_q (n) &=& \frac{(-1)^{q - 1}}{[q n - (q - 1)]^{3/2}} \prod_{s = 0}^{q - 1} [s n - (s - 1)] ,
\end{eqnarray}
which are characteristic coefficients of individual NLA processes of order $n=m$ and $l$, being $q = 1,2,...$ a positive integer.
Additionally, an infinite number of coupled coefficients arise in (\ref{eq17}), some of those of the lowest order being
\begin{subequations}
\begin{eqnarray}
 C_{1 1} (m,l) &=& - \frac{m + l}{ (m + l - 1)^{3/2}}                                          \\
 C_{2 1} (m,l) &=& \frac{3 m^2 + 2 m l + l^2 - 2 m - l}{ (2 m + l - 2)^{3/2}}                  \\
 C_{3 1} (m,l) &=& - \frac{(4 m^2 + m l + l^2 - 3 m - l) (3 m + l - 2)}{ (3 m + l - 3)^{3/2}}  \\
 C_{2 2} (m,l) &=& - \frac{(7 m^2 + 4 m l + 7 l^2 - 6 m - 6 l) (m + l - 1)}{ (2 m + 2 l - 3)^{3/2}}
\end{eqnarray}
\end{subequations}
The property of symmetry $C_{p q} (m,l) = C_{q p} (l,m)$ allows a faster computation of different coefficients. 
As inferred in a previous discussion, as long as the peak depths $\Delta_n$ approach the unity, accurate estimation of $A_{max}$ is expected to involve running terms of increasing index $S = p + q$ in the summation.

%Begining at the point that experimental data tracing the $Z$-scan signature may be processed in order to estimate $m$ and $\Delta_m$, the peak depth (apart of $l$).
%Its value may be determined directly from the measured absorptance at focus $A_{max}$ using either Eq.~(\ref{eq17}) or ultimately the numerical calculations shown in Fig.~\ref{fig07}.

\section{\label{sec05} Conclusions}

We have performed a geometrical analysis of the open aperture $Z$-scan trace when multiphoton absorption of different nature arise, simultaneously, in the material under inspection.
Under appropriate conditions, a simple analytical expression may be found allowing an accurate description of the beam shaping induced by absorption.
Parametrization of the nonlinear beam propagation is formulated in terms of a series expansion providing an analytical representation of the absorptance produced in the nonlinear medium as experimentally recorded in the traces.
%This formalizes mathematically the well-known fact that higher-order nonlinear absorption modulates strongly the sample transmittance at the valley of the $Z$-scan signature, as long as the trace tails have a characteristic decay revealing the lower-order nonlinearity.
%
%In conclusion, we have developed a procedure to determine the values of NLA coeffients from $Z$-scan measurements when two independent absorptions are experienced in the sample.
Unlike fitting $Z$-scan signatures directly, we propose the evaluation of asymptotes arisen when absorptance are represented versus the width of the Gaussian beam at different scanning tracks.
%This analysis is performed assuming Gaussian profiles of the probe beam in space and time domains.
%; however, a generalization including different patterns and pulse forms may be given straightforwardly.
A fundamental drawback of the procedure is the extreme sensitivity of data measurement required to draw the estimates out for the coefficient of the highest order.
To overcome this major inconvenient, the peak value of the absorptance is suggested to play on the determination of the remaining NLA coefficient.

\begin{acknowledgments}

The author is indebted to Florencio E. Hern\'andez for stimulating conversations.
This research was funded by the Generalitat Valenciana under the project GVPRE/2008/005.
The author also acknowledges financial support from Ministerio de Ciencia e Innovaci\'on.
%C. J. Zapata-Rodr\'{\i}guez's e-mail address is carlos.zapata@uv.es

\end{acknowledgments}

%\bibliographystyle{osajnl}
%\bibliography{Bibliography}

\end{document}